# Integrated Multiport Bidirectional DC-DC Converter for HEV/FCV Applications

Bang Le-Huy Nguyen[†§], Honnyong Cha[‡], Tuyen Vu[†], Thai-Thanh Nguyen[†],
[†]*ECE Department, Clarkson University*, Potsdam, NY, USA
[‡]*School of Energy Engineering, Kyungpook National University*, Daegu, Korea
[§]*National Renewable Energy Research Lab*, Golden, CO, USA
nguyenbl@clarkson.edu, chahonny@knu.ac.kr, tvu@clarkson.edu, tnguyen@clarkson.edu

*Abstract*— **This paper proposes a novel integrated multiport bidirectional dc-dc converter to interface the battery, the ultra-capacitor, the fuel cell, or other energy sources with the dc-link capacitor of the hybrid energy systems such as the hybrid electric vehicle (HEV) and fuel cell vehicle (FCV) applications. The proposed converter can be applied to the distributed generation systems which include local energy sources, storage, and loads. It can perform both buck and boost functions with fewer switches. In addition, it is extendable when more inputs and/or outputs are required. The operating principle and control strategy of the proposed converter will be analyzed in detail. For verification, simulation, and experimental results of the four utilized operating modes of an HEV/FCV are provided.**

*Keywords*— **DC-DC power converter, fuel cell vehicle (FCV), hybrid electric vehicle (HEV), multi-port converter.**

## I. INTRODUCTION

Electric vehicles have been received increasing attention in recent years and considered as an alternative to the conventional internal combustion engine vehicle. The HEV/FCV with a fuel cell or other energy sources are preferred when the driving range is long and for where the charge station is not well distributed. The dc-dc converters play a major role in the energy transfer between the fuel cell, the drives of the propulsion motor, and the battery.

In addition, to enhance the lifetime of the fuel cell and the battery, ultra-capacitors with a higher number of charge/discharge cycles are included [1]. The ultra-capacitor helps the fuel cell and the battery operate at the constant transferred power mode by storing the excess energy in a low-power mode or the regenerative braking period and providing more energy in high-power mode during acceleration or uphill driving. Thus, it also increases the efficiency of the whole system [2]. Nonetheless, one more converter module is required to regulate the in-out power of the ultra-capacitor.

To effectively manage the power flow between the above elements, all converters should be regulated as a power electronic interface unit [3]. Accompanying to this, the multiport bidirectional power converter has been attracting research interests owing to lower cost, higher

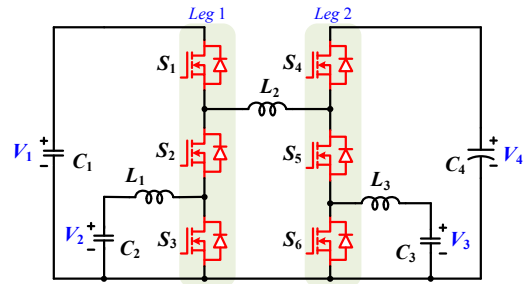

Fig. 1. Proposed integrated multiport bidirectional DC-DC converter

power density, and better power flow control, and faster dynamic response.

There are many bidirectional multiport dc-dc power converters are introduced for HEV/FCV and distributed generation systems in the literature [4]–[12]. In [4], a bidirectional converter using a couple of inductors was proposed. In this topology, the ultra-capacitor and the battery are cascaded and they are connected in parallel with the dc-link voltage. Thus, their total voltage is always the same as the dc-link voltage. In addition, the FC feeds to the dc-link just flowing through a diode. These issues limit the control ability of the converter. In [5], the Z-source inverter (ZSI) for HEV with the battery connected to one of the Z-network capacitors is proposed. Although this concept can regulate the FC output power and the state of charge of the battery at the same time. The battery voltage is dependent on the boost factor of ZSI. Also, there is no ultra-capacitor. The paper [6] introduced a multi-output synchronous buck converter with a smaller number of switches. However, the buck outputs do not have the same ground and their total voltage is also limited. In [7], a multi-input bidirectional DC-DC converter that can operate in both buck and boost modes with flexible power flow management is derived by using the two-switch legs. This converter is the combination of two conventional non-inverting buck-boost dc-dc converters by sharing a leg. However, by connecting the battery and a switch in series, the battery current might be discontinuous which degrade its lifetime.

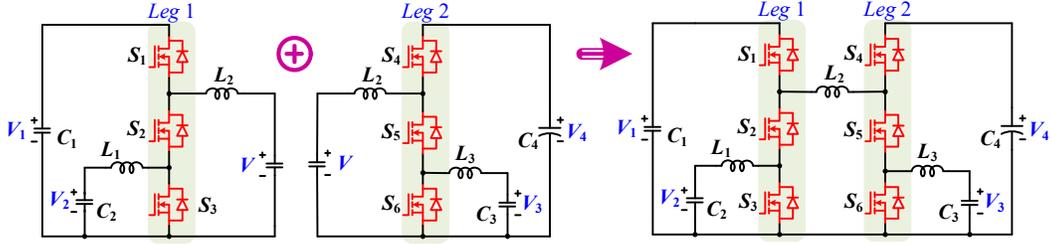

Fig. 2. The derivation of the proposed converter.

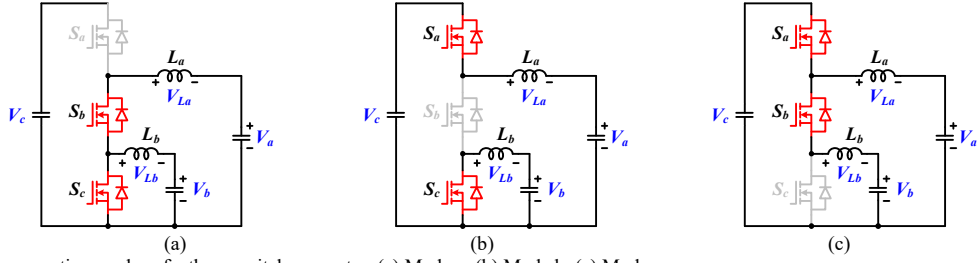

(a)　　　　　　　　　(b)　　　　　　　　　(c)

Fig. 3. The operating modes of a three-switch converter. (a) Mode a. (b) Mode b. (c) Mode c.

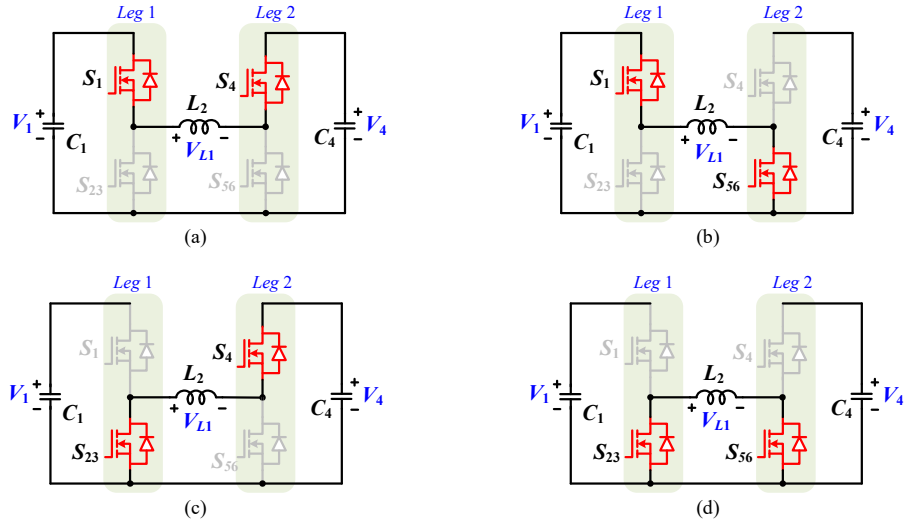

Fig. 4. The operating modes of the non-inverting buck-boost converter. (a) Mode 1. (b) Mode 2. (c) Mode 3. (d) Mode 4.

This paper proposes a novel integrated multiport dc-dc converter that can provide the bidirectional power flow between any four ports with both bucks and boost functions as shown in Fig. 1. All the ports have the same ground. The power transfer between any two can be regulated independently at the same time. The voltage port $V_1$ can be greater or smaller than the voltage port $V_4$. The currents at ports 2 and 3 are continuous, then these ports are suitable for the battery. It is possible to extend to more ports if required. The design and implementation are simple, the control is flexible.

## II. OPERATION ANALYSIS

The proposed converter circuit is derived from the combination of two bidirectional three-switch converters as illustrated in Fig. 2. This is similar to the derivation of the well-known non-inverting buck-boost dc-dc converter [13]–[15]. As a result, the proposed converter can perform both bidirectional bucks and boost functions from port $V_1$ to port $V_4$ and vice versa.

For simplicity, each three-switch leg is considered independently in the operation. Firstly, the operating modes of a three-switch converter are analyzed. Then, combing with the operating principle of the non-inverting buck-boost converter, the voltage gain between the ports of the proposed converter can be achieved.

### A. Three-Switch Converter

With the assumption that the inductor currents are continuous, only one switch is turned off, while the other two are turned on at any time. There are three modes with three duty times in the operation defined as follows. The equivalent circuits of the modes of a three-switch converter are illustrated in Fig. 3.

*1) Mode a* (duty ratio $D_a$): the switch $S_a$ is turned off, while the switches $S_b$ and $S_c$ are turned on. The voltage across the inductors $L_a$, $L_b$ are:

$$V_{La} = -V_a, \quad V_{Lb} = -V_b \qquad (1)$$

2) *Mode b (duty ratio $D_b$):* the switch $S_b$ is turned off, while the switches $S_a$ and $S_c$ are turned on. The voltage across the inductors $L_a$, $L_b$ are expressed as:

$$V_{La} = V_c - V_a, \quad V_{Lb} = -V_b \qquad (2)$$

3) *Mode c (duty ratio $D_c$):* the switch $S_c$ is turned off, while the switch $S_a$ and $S_b$ are turned on. The voltage across on the inductors $L_a$, $L_b$ are determined as:

$$V_{La} = V_c - V_a, \quad V_{Lb} = V_c - V_b \qquad (3)$$

Where,

$$D_a + D_b + D_c = 1 \qquad (4)$$

Applying the flux balance condition for inductor $L_a$ and $L_b$ we have the below equations, respectively.

$$-V_a D_a + (D_b + D_c)(V_c - V_a) = 0 \qquad (5)$$

$$-V_b(D_a + D_b) + D_c(V_c - V_b) = 0 \qquad (6)$$

The relationship between the port voltage can be derived as:

$$V_a = (D_b + D_c)V_c, \quad V_b = D_c V_c \qquad (7)$$

As can be seen, the port voltage $V_b$ is independent of $V_a$, it only depends on the port voltage $V_c$ and the duty ratio of mode c ($D_c$).

Similarly, we can define the duty ratios ($D_1$–$D_6$) are the proportion time when the switches ($S_1$–$S_6$) are turned off, respectively. According to the above analysis, we have:

$$V_2 = D_3 V_1, \quad V_3 = D_6 V_4 \qquad (8)$$

The relationship between the voltage ports $V_1$ and $V_4$ can be achieved based on the operation of the non-inverting buck-boost converter as the below analysis.

### B. Non-inverting Buck-Boost Converter

For simplicity, the ports $V_2$ and $V_3$ are neglected. The lower switches ($S_2$, $S_3$) and ($S_5$, $S_6$) are combined into the switches $S_{23}$ and $S_{56}$, respectively. The proposed converter now can be considered as a conventional non-inverting buck-boost converter with four operating modes as shown in Fig. 4.

With $D_1$ and $D_4$ representing the duty cycles when the switches $S_1$ and $S_2$ turned off respectively, the relationship between the voltage ports $V_1$ and $V_2$ can be expressed [13]

$$(1 - D_1)V_1 = (1 - D_4)V_4 \qquad (9)$$

When $D_1 = 0$, the switch $S_1$ is always turned on, the inductor $L_2$ is connected directly to $C_1$. The proposed converter operates only in mode 1 and mode 2. It becomes a bidirectional boost converter from $V_1$ to $V_4$. The equation (9) becomes $V_4 = V_1/(1-D_4)$, ($V_1 \leq V_4$).

Conversely, when $D_4 = 0$, the switch $S_4$ is always turned on, the inductor $L_2$ is connected directly to $C_4$. The proposed converter operates only in mode 1 and mode 3. It becomes a bidirectional buck converter from $V_1$ to $V_4$. The equation (9) becomes $V_4 = V_1(1-D_1)$, ($V_1 \geq V_4$).

Conventionally, the above control scheme is used when $V_1$ is larger or smaller than $V_4$. However, when $V_1$ fluctuates within a small boundary around $V_4$, this scheme is difficult to perform. Since some practical gate driver circuits with the bootstrap technique cannot generate duty ratios that are smaller than 5 % or larger than 95 %.

A simple solution can be applied to solve this issue. When the $V_1$ and $V_4$ values are near together, the duty ratio $D_1$ is kept at a fixed value (0.2, for example). Thus, the duty ratio $D_2$ can be controlled around $D_1$ so that the equation (9) is satisfied.

Summarily, the power transfers between the ports $V_2$ and $V_1$, $V_3$ and $V_4$ can be regulated by controlling the duty ratios $D_3$ and $D_6$, respectively, as given in equation (8). Whereas the power transfer between the ports $V_1$ and $V_4$ can be performed by controlling the duty ratios $D_1$ and $D_4$ as given in equation (9).

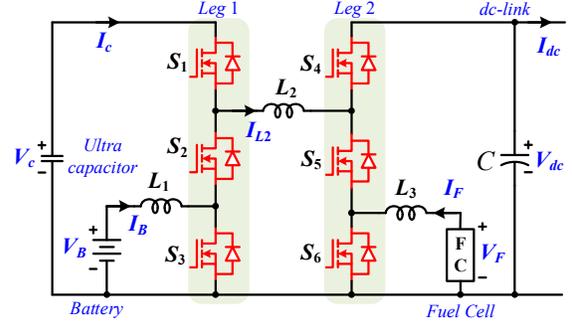

Fig. 5. The configuration of the proposed converter for HEV/FCV applications

## III. CONFIGURATION AND CONTROL MODES OF PROPOSED CONVERTER FOR HEV/FCV

The configuration of the proposed converter for HEV/FCV applications is illustrated in Fig. 5. In the first leg, the ports $V_1$, $V_2$ are connected to the ultra-capacitor and the battery, respectively. While, in the second leg, the ports $V_3$ and $V_4$ are employed to interface the fuel cell and the dc-link which feeds to the propulsion motor's drive.

The control of the proposed converter is analyzed under the four utilized operating modes [5] of an HEV/FCV application as below. The simulation results are also provided to describe the voltage and current waveforms.

For simplicity, in the simulation, the ports are connected to the voltage sources and the resistive loads when receiving and feeding the power, respectively. The battery voltage is $V_B = 25$ V and the fuel cell voltage is $V_F = 35$ V. The dc-link voltage is regulated at $V_{dc}$=100 V. The ultra-capacitor can be regulated at the voltage higher or lower than $V_{dc}$. The inductance value of $L_1$–$L_3$ is 0.72 mH. The switching frequency is 50 kHz.

### A. Medium Power Mode

In this mode, the propulsion motor only receives the energy from the fuel cell. The ultra-capacitor is employed to keep the fuel operate under constant power when the required power at the dc-link fluctuates. The battery might be charged by the power from the fuel cell if it is low in the state of charge.

Fig. 6 shows the voltages and currents waveforms of the converter. Where the output current at the dc-link is

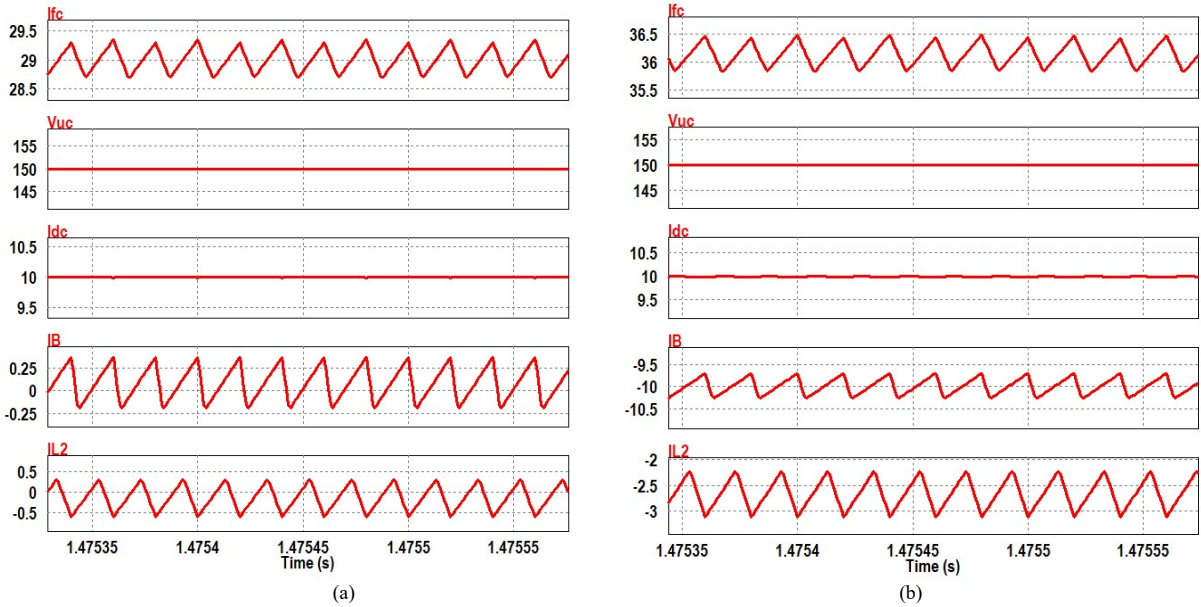

(a) (b)
Fig. 6. The simulated waveforms of the dc-link, ultra-capacitor voltages, and the dc-link, battery, inductor $L_2$ currents (from top to bottom) at the medium-power mode when the battery is remained (a) and charged (b).

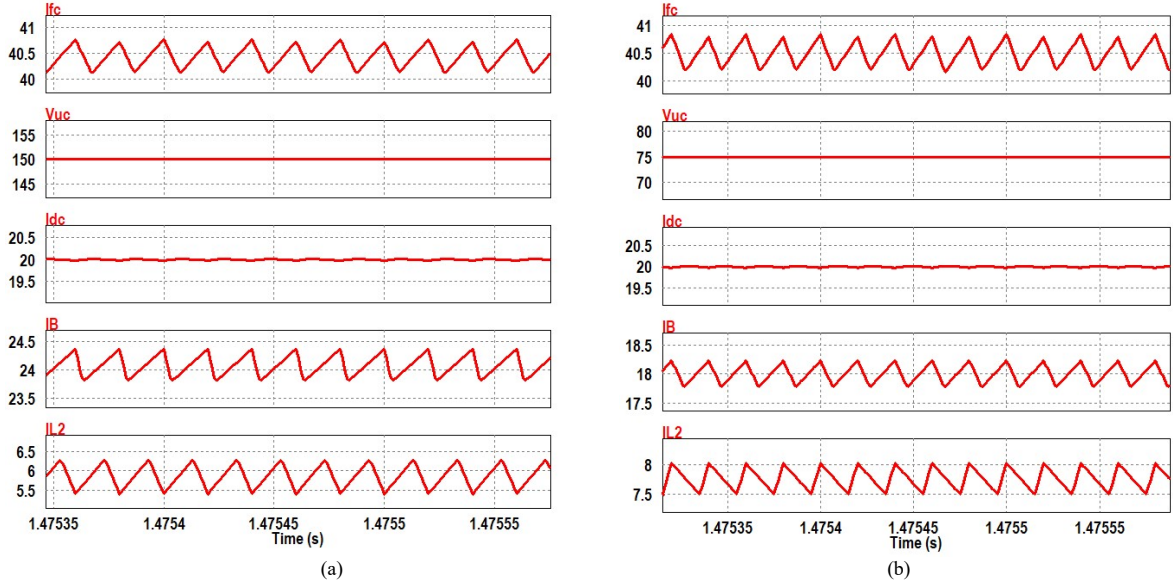

(a) (b)
Fig. 7. The simulated waveforms of the fuel cell current, the ultra-capacitor voltages, and the dc-link, battery, inductor $L_2$ currents (from top to bottom) at the high-power mode when the ultra-capacitor voltage is remained at 150 V (a) and discharged at 75 V (b).

10 A. In Fig. 6(a), the fuel cell feeds all the required power. The ultra-capacitor voltage is regulated at 150 V. Averagely, it does not absorb or eject any power. The battery's state of charge is assumed at high. Thus, the battery current is kept at zero. Since the three-switch leg 1 does not transfer any power to leg 2, the inductor current $I_{L2}$ also remains around zero.

Fig. 6(b) shows the waveforms when the state of charge of the battery is low. Although the output current remains the same, the fuel cell should eject more current to charge the battery. The battery is charged with 10 A. The inductor current $I_{L2}$ reflects the charged power transferred from leg 1 to leg 2.

*B. High Power Mode*

When the vehicle accelerates or drives uphill, it needs more energy. Thus, the dc-link needs to receive power from both the FC and the battery. The ultra-capacitor can be charged, discharged, or remained. This can be decided based on the fluctuation of the required power. Simulation results for this mode are shown in Fig. 7.

Assumed that the dc-link current is 20 A, the required power is 2 kW. As shown in Fig. 7(a), the fuel cell ejects a power of about 1.4 kW by the fuel cell current of 40 A. While the battery provides the remained power of 0.6 kW by the battery current of 24 A. Where the ultra-capacitor voltage remains at 150 V.

In Fig. 7(b), the fuel cell current is kept the same.

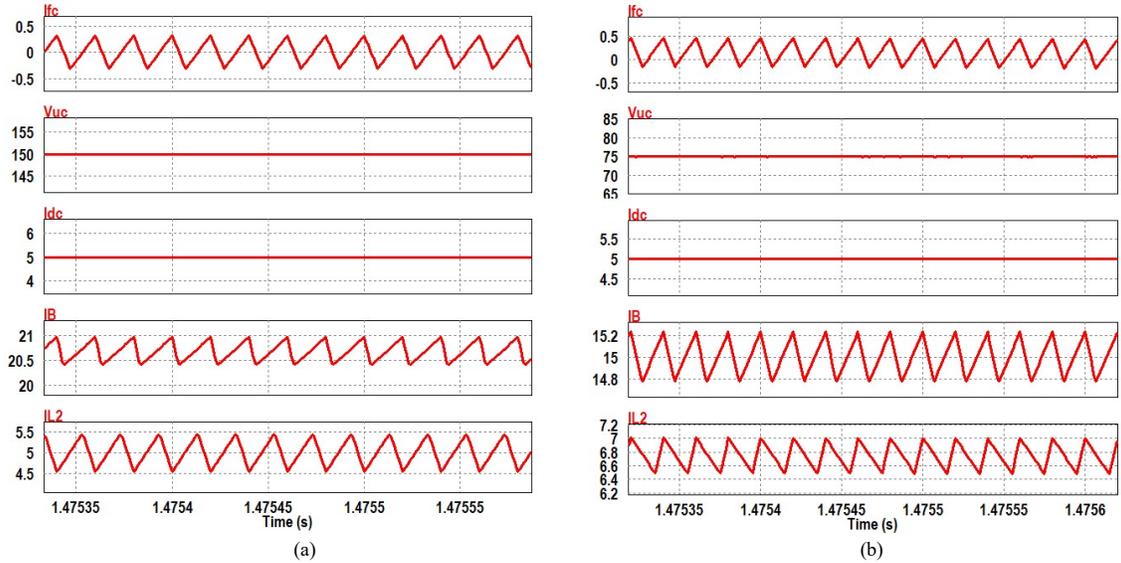
Fig. 8. The simulated waveforms of the dc-link, ultra-capacitor voltages, and the dc-link, battery, inductor $L_2$ currents (from top to bottom) at the low-power mode when the ultra-capacitor voltage is remained at 150 V (a) and discharged at 75 V (b).

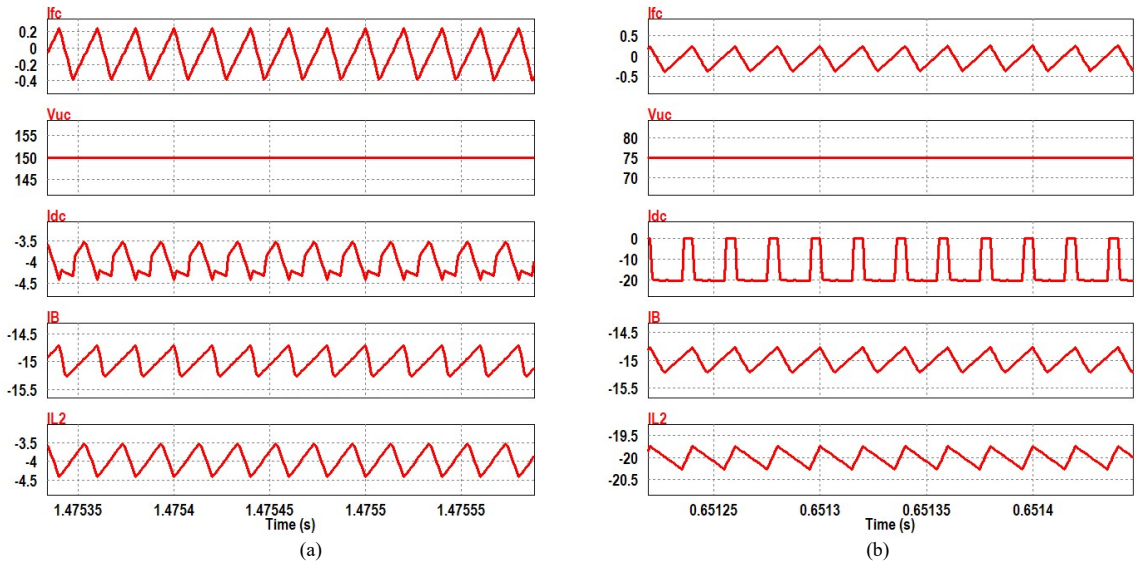
Fig. 9. The simulated waveforms of the fuel cell current, the ultra-capacitor voltages, and the dc-link, battery, inductor $L_2$ currents (from top to bottom) at the regenerative-braking mode when the ultra-capacitor voltage is remained at 150 V (a) and charged at 75 V (b).

However, the battery current is reduced to 18 A. Thus, the remained power is provided by the ultra-capacitor which has a voltage of 75 V at this time.

### C. Low-Power Mode

When the vehicle is moving flatly and at a constant speed. The converter system can operate under low-power mode. When the fuel cell stack is turned off due to high power losses on the accompanying systems such as the air compressor. Therefore, only the battery and the ultra-capacitor can provide the power. The simulation results for this mode are shown in Fig. 8.

Where the output current is assumed to reduce to 5 A. The fuel cell current is zero. The battery supplies to the dc-link by a current of about 21 A, when the ultra-capacitor remains at 150 V as shown in Fig. 8(a). The battery current is reduced to about 15 A when the ultra-capacitor also provides the power to output as shown in Fig. 8(b). The ultra-capacitor voltage now is 75 V.

### D. Regenerative Braking Mode

When the vehicle goes downhill or brakes, the energy can be transferred back to the converter and then stored in the ultra-capacitor or the battery. Simulation results for this mode are shown in Fig. 9

Where the fuel cell current is zero, the output current is negative. Fig. 9(a) shows the charging of the battery with the current of 15 A. The ultra-capacitor remains at 150 V.

When the ultra-capacitor voltage is low, it also can be charged. As shown in Fig. 9(b), the reverse output current is larger, however, the charging current of the battery is kept the same. Thus, the remained energy is stored in the ultra-capacitor.

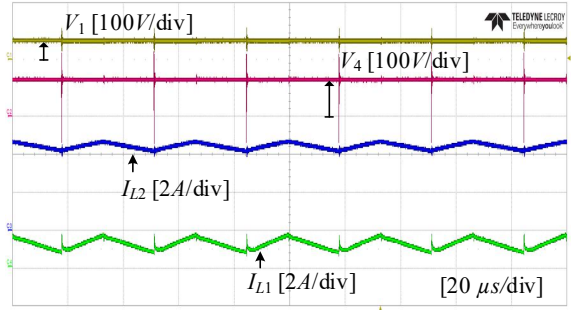

Fig. 10. The experimental waveforms for the proposed converter when the port $V_2$ is the input source.

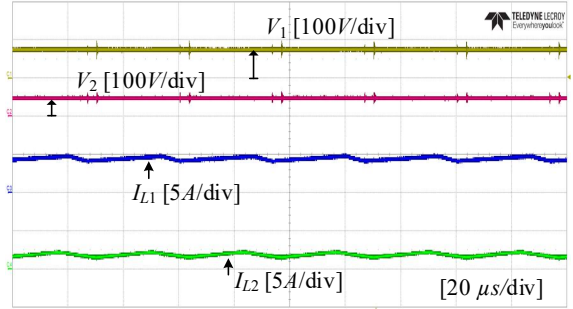

Fig. 11. The experimental waveforms for the proposed converter when the port $V_4$ is the input source.

## IV. EXPERIMENTAL RESULTS

The hardware prototype of the proposed converter is built and tested. The switches are MOSFETs 47N60CFD of Infineon. The inductor values of $L_1$–$L_3$ are 0.72 mH. The switching frequency is 30 kHz. The capacitor filter is sufficient to keep voltages stable.

Firstly, port $V_2$ is connected to a voltage source of 25 V, whereas ports $V_1$ and $V_4$ are connected to the resistive loads of 100 Ω. The port $V_3$ is not employed. Fig. 10 shows the experimental waveforms of the port voltages $V_1$ and $V_4$ and the inductor currents of $L_2$ and $L_1$. In this test, the input source $V_2$ is charged the inductor $L_2$ to push the power to the port $V_1$. Then, port $V_1$ can perform a boost function by the inductor $L_2$ to transfer the power to port $V_4$.

The second test is implemented with the input source 100 V connected to port $V_4$. Then, the proposed converter performs a buck function to transfer the power to port $V_1$ through the inductor $L_1$. Then, the port voltage $V_1$ is continuously bucked to the port voltage $V_2$ by the inductor $L_2$. Fig. 11 shows the experimental waveforms of the port voltages and inductor currents in this test.

## V. CONCLUSIONS

A novel integrated multi-port dc-dc converter that can provide bidirectional power transfer between any two ports is proposed in this paper. The proposed converter can be employed for interfacing sources, loads, storage devices in distributed generation systems or dc micro/nano-grids including the HEV/FC power systems. The operational principle of the proposed converter is analyzed in detail. The configuration of the proposed converter with the battery, the ultra-capacitor, the fuel-cell interfacing with the dc-link for an HEV is proposed. The simulation results under four utilized operation modes of the HEV are investigated. The experimental results are given to prove the theoretical analysis.

## REFERENCES

[1] E. Schaltz, A. Khaligh, and P. O. Rasmussen, "Influence of battery/ultracapacitor energy-storage sizing on battery lifetime in a fuel cell hybrid electric vehicle," *IEEE Trans. Veh. Technol.*, vol. 58, no. 8, pp. 3882–3891, Oct. 2009.

[2] A. Khaligh and Z. Li, "Battery, ultracapacitor, fuel cell, hybrid energy storage systems for electric, hybrid electric, fuel cell, plug-in hybrid electric vehicles: State of the art," *IEEE Trans. Veh. Technol.*, vol. 59, no. 6, pp. 2806–2814, Jul. 2010.

[3] M. McDonough, "Integration of inductively coupled power transfer and hybrid energy storage system: a multiport power electronics interface for battery-powered electric vehicles," *IEEE Trans. Power Electron.*, vol. 30, no. 11, pp. 6423-6433, Nov. 2015.

[4] W.-S. Liu, J.-F. Chen, T.-J. Liang, R.-L. Lin, and C.-H. Liu, "Analysis, design, control of bidirectional cascoded configuration for a fuel cell hybrid power system," *IEEE Trans. Power Electron.*, vol. 25, no. 6, pp. 1565–1575, Jun. 2010.

[5] F. Z. Peng, M. Shen, and K. Holland, "Application of Z-source inverter for traction drive of fuel cell- battery hybrid electric vehicles," *IEEE Trans. Power Electron.*, vol. 22, no. 3, pp. 1054–1061, May 2007.

[6] G. Chen, Y. Deng, J. Dong, Y. Hu, L. Jiang, and X. He, "Integrated multiple-output synchronous buck converter for electric vehicle power supply," *IEEE Trans. Veh. Technol.*, vol. 66, no. 7, pp. 5752-5761, Jul. 2017.

[7] A. Hintz, U. R. Prasanna, and K. Rajashekara, "Novel modular multiple-input bidirectional DC–DC power converter (MIPC) for HEV/FCV application," *IEEE Trans. Ind. Electron.*, vol. 62, no. 5, pp. 3163-3172, May 2015.

[8] W. Jiang and B. Fahimi, "Multiport power electronic interface—concept modeling and design," *IEEE Trans. Power Electron.*, vol. 26, no. 7, pp. 1890-1900, Jul. 2011.

[9] P. Shamsi and B. Fahimi, "Dynamic behavior of multiport power electronic interface under source/load disturbances," *IEEE Trans. Ind. Electron.*, vol. 60, no. 10, pp. 4500-4511, Oct. 2013.

[10] M. McDonough, "Integration of inductively coupled power transfer and hybrid energy storage system: a multiport power electronics interface for battery-powered electric vehicles," *IEEE Trans. Power Electron.*, vol. 30, no. 11, pp. 6423-6433, Nov. 2015.

[11] Y. Li, X. Ruan, D. Yang, F. Liu, and C. K. Tse, "Synthesis of multiple-input DC/DC converters," *IEEE Trans. Power Electron.*, vol. 25, no. 9, pp. 2372-2385, Sep. 2010.

[12] Y. C. Liu and Y. M. Chen, "A systematic approach to synthesizing multi-input DC–DC converters," *IEEE Trans. Power Electron.*, vol. 24, no. 1, pp. 116-127, Jan. 2009.

[13] N. Zhang, S. Batternally, K. C. Lim, K. W. See and F. Han, "Analysis of the non-inverting buck-boost converter with four-mode control method," *IECON 2017 - 43rd Annual Conf. of the IEEE Ind. Electron. Soc.*, Beijing, 2017, pp. 876-881.

[14] C. Yao, X. Ruan, W. Cao, and P. Chen, "A two-mode control scheme with input voltage feed-forward for two-switch buck-boost dc-dc converter," *IEEE Trans. Power Electron.*, vol. 29, no. 4, pp. 2037–2048, Apr. 2014.

[15] X. E. Hong, J. F. Wu, and C. L. Wei, "98.1%-Efficiency Hysteretic-Current-Mode Noninverting Buck–Boost DC-DC Converter with Smooth Mode Transition," *IEEE Trans. Power Electron.*, vol. 32, no. 3, pp. 2008-2017, Mar. 2017.